# Quantum dot micropillar cavities with quality factors exceeding 250,000

C. Schneider[1], P. Gold[1], S. Reitzenstein[1,2], S. Höfling[1,3] and M. Kamp[1]

[1]Technische Physik, Physikalisches Institut and Wilhelm Conrad Röntgen-Research Center for Complex Material Systems, Universität Würzburg, Am Hubland, D-97074, Würzburg, Germany
[2]Present address: Institut für Festkörperphysik, Technische Universität Berlin, Hardenbergstraße 36, D-10623 Berlin, Germany
[2]SUPA, School of Physics and Astronomy, University of St Andrews, St Andrews, KY16 9SS, United Kingdom
[*]Christian.schneider@physik.uni-wuerzburg.de

**Abstract:** We report on the spectroscopic investigation of quantum dot - micropillar cavities with unprecedented quality factors. We observe a pronounced dependency of the quality factor on the measurement scheme, and find that significantly larger quality factors can be extracted in photoreflectance compared to photoluminescence measurements. While the photoluminescence spectra of the microcavity resonances feature a Lorentzian lineshape and Q-factors up to 184,000 ($\pm$10,000), the reflectance spectra have a Fano-shaped asymmetry and feature significantly higher Q-factors in excess of 250,000 resulting from a full saturation of the embedded emitters. The very high quality factors in our cavities promote strong light-matter coupling with visibilities exceeding 0.5 for a single QD coupled to the cavity mode.

1. Introduction

Dielectric distributed Bragg reflectors (DBRs) monolithically grown by means of molecular beam epitaxy or metal-organic vapor phase epitaxy are key building blocks for tailoring light confinement in nanophotonics devices. They are widely utilized in state-of-the-art vertically emitting microcavity lasers [1], optical filters, spin-photon interfaces [2] and they might also be useful for increasing the efficiency of solar cells [3,4]. However, the very high quality factors that can be provided by DBR-based microcavity structures also explains their important role in fundamental semiconductor optics, in particular in the research field of light-matter interaction in semiconductors[5]. Here, DBRs are commonly sandwiching an optical defect layer which breaks the translation symmetry of the system. This layer usually contains the active material such as semiconductor quantum dots. In such a microcavity, photons can be 'stored' for a given number of roundtrips, until they leak out of the cavity. This storage time is proportional to the Q-factor of the microcavity, which is a unit-less quantity describing the cavity's capability to store optical energy. In systems with integrated quantum emitters, predominantly the parameters of the cavity determine the interaction regime: In the weak light-matter coupling regime, the radiative lifetime of the emitter is altered by the presence of the cavity via Fermi's golden rule [6]. In contrast, if the light-matter interaction strength exceeds the cavity loss channels, the coherent regime of strong coupling is reached [7-9]. Both regimes have fundamental importance in the design of semiconductor devices with integrated quantum emitters (quantum dots) of the 'next generation' of photonic devices, such as efficient sources of single photons on demand[10-12], sources of entangled photon pairs[13], and sources of coherently generated and emitted single photons as demonstrated for atoms in optical cavities[14] . High-Q microcavities with embedded quantum dots also play a major role in the development of key building blocks for solid state quantum repeaters, in particular for developing spin photon interfaces [15,16]. This motivates the need of high quality microcavities. Q-factors of state-of-the-art quantum dot- cavities as high as 165,000 were reported in photoluminescence on devices with diameters of 4 µm [17] and in excess of 200,000 for a pillar with a diameter of 7.3 µm in photoreflectance[18]. We report here Q-factors exceeding 250,000 for a 6 µm diameter micropillar cavity measured in reflectance defining the state of the art and we observe a characteristic feature in the reflectance spectra being associated with a Fano-resonance. The Q-factor of a microcavity is commonly determined by extracting the spectral width of its photonic resonance. However, depending on the type of microcavity and the measurement technique, also the shape of the resonance can significantly alter: The photoluminescence spectra of semiconductor microcavties with embedded active material, and in the absence of strong photonic disorder, typically feature a Lorentzian lineshape. The width of the resonance is then directly determined by the photon lifetime in the resonator. Recently, reflectance measurements of photonic crystal nanoresonators have revealed strong Fano-features in the lineshape as a result of the interference between an effective two level photonic system and continuum modes [20,21]. Here, we carry out a comparative investigation of micropillar cavities with ultra-high quality factors via micro-photoluminescence (µPL) and micro-photoreflectance (µPR) with respect to their quality factor as well as the shape of the resonance.

2. Experiment and Discussion

The structure under consideration for the following in-depth study consists of a microcavity with 36 (32) AlAs/GaAs layer pairs in the bottom (top) Bragg mirror. Each mirror segment was designed to match the $\frac{\lambda}{4n}$ Bragg condition with corresponding thicknesses of 68 nm for the GaAs layer and 81.5 nm of the AlAs layers. The intrinsic GaAs-$\lambda$-cavity (nominal thickness~ 272 nm) contains a single layer of low density, $In_{0.30}Ga_{0.70}As$ quantum dots (QDs) with a nominal area density of ~2-4 $10^8$ 1/cm$^2$ and a large oscillator strength. Micropillars were defined by electron beam lithography and etched into the layer structure by electron-cyclotron-resonance reactive-ion-etching. The etch technique has been optimized to achieve a maximum aspect ratio, which is reflected by the highly vertical sidewalls with minimum roughness, as can be seen from the scanning electron microscope image in Fig. 1a). A detailed analysis of the sidewall morphology, as well as more details concerning the sample fabrication technique can be found e.g. in [17,19] and the references therein.

In order to study the Q-factor of this cavity, we exploit two complementary techniques: In µPR, the Q-factor can be determined by measuring the width of the resonance dip. However, we can also use the integrated QDs as an internal light source, allowing us to probe the Q-factor in the µPL configuration. This approach is usually chosen to determine the Q-factor of such cavities due to its simplicity. However, this technique is only applicable in active structures and if a sufficiently high number of QDs are spectrally located close, within about 10 nm, to the cavity resonance to facilitate its illumination via non-resonant QD-cavity coupling effects [12]. A schematic drawing of the experimental setup for both configurations is shown in Fig. 1b). The slight ellipticity, which is present in our micropillars, results in a linear polarization splitting of the fundamental cavity mode, which ranges between 0- 50 µeV, i.e on the order of the microcavity linewidth [17]. In the following, only one of these two resonance is studied by introducing a linear polarizer in the beampath.

Fig. 2a) shows a typical µPL spectrum of a micropillar cavity with a diameter of 8 µm at 14 K. The pillar is excited by a frequency-doubled Nd:YAG laser focused to a spot-size of 4 µm operated in continuous wave mode at a wavelength of 532 nm. The signal from the sample is dispersed in a double monochromator with a telescope attached to the exit slit. The extracted emission line can be reproduced by a Voigt profile which convolves the Lorentzian spectral line shape from the cavity emission with the spectral response from the spectrometer with a resolution of 6.2 µeV. The Lorentzian contribution, which is directly related to the photon lifetime in the optical resonator via Fourier transformation yields a cavity linewidth as small as 5.14 pm (7.15 µeV±0.5 µeV), which directly converts into a Q-factor of 184,000±10,000. This value reflects the high quality of the QD micropillar fabrication process, and is in good agreement with previous photoluminescence studies of micropillar cavities with such a large number of dielectric layers [17]. The µPR spectrum of the same pillar is shown in Fig. 2b). A tunable diode laser with a linewidth of 100 kHz with an optical (output) power on the order of ~ 20 µW, focused to a slightly larger spot size of ~7 µm, was scanned across the resonance. The wavelength of the laser was accurately monitored by a wavelength meter, and the reflected power was recorded by a Silicon photodetector. Noteworthy, and in stark contrast to the µPL signal, the lineshape of the reflection dip features a Fano-lineshape: We explain this peculiarity by a resonant and a non-resonant contribution of reflected light to the signal. The resonant, quasi zero-dimensional scattering channel is represented by the optical microcavity mode, while the incoming laser beam additionally excites higher lateral modes and continuum modes due to an imperfect mode matching. This leads to an interference effect, which is manifested by the Fano-lineshape of the reflection spectrum. While similar effects caused by a mode mismatch between the incoming laser beam and the resonator have been observed in photonic crystal nanocavities [20,21], we note that a fully microscopic understanding of the origin of the Fano-effect in our device requires advanced numerical simulations, which is beyond the scope of this paper. We can reproduce the reflection-spectrum by the formula:

$$F(\omega) = R_0 + A_0 \frac{(q+2(\omega-\omega_0)/\gamma_c)^2}{1+(2(\omega-\omega_0)/\gamma_c)^2} \qquad (1)$$

Here, $R_0$ and $A_0$ are constants for reflectivity offset and amplitude, $\omega_0$ is the frequency and $\gamma_c$ the linewidth of the cavity mode. The Fano parameter q is given by the ratio of the resonantly and non-resonantly scattered light.

By applying this model to our system, we can extract the linewidth of the cavity mode $\gamma_c = 1.2\ Ghz$ ($4.9\ \mu eV$) and hence a cavity Q-factor as high as 268,000, which compares favourable with the current state-of-the-art for micropillar cavities [17,18]. Noteworthy, such a high Q-factor corresponds to a photon storage time of ~ 130 ps in our micropillar resonator. Surprisingly, we observe a significantly higher Q-factor in the reflection technique as compared to the photoluminescence case. This systematic deviation between the extracted values is shown in Fig. 2c) where we plot the extracted Q-factors as a function of the pillar diameter both for the PL and the reflection case. We can accurately reproduce the measured diameter dependency of the Q-factors by a model taking into account several photon loss channels according to the formula [22,23]:

$$\frac{1}{Q} = \frac{1}{Q_{int}} + \frac{1}{Q_{scat}} + \frac{1}{Q_{abs}}. \tag{2}$$

The first term $\frac{1}{Q_{int}}$ describes the intrinsic losses of the microcavity. It is determined by photon leakage through the mirrors without taking into account any material absorption. In our sample, the theoretical planar Q-factor has a value of $2.6 \cdot 10^6$ and consequently other photon loss channels dominate the system. As discussed in Refs.[22,24] for the case of micropillar cavities, this intrinsic loss term successively increases towards smaller diameters due to the spectral mode shift towards the edge of the stop-band. The second term in Eq. (2) describes sidewall losses specifically in micropillars. The term takes into account a finite intensity of the electromagnetic field at the sidewall of the micropillar relative to its circumference and can be approximated via $\frac{1}{Q_{scat}} = \frac{\kappa J_0^2(kR)}{R}$ [22]. Here, R is the radius of the micropillar and $J_0^2(kR)$, which is the Bessel function of $0^{th}$ order, is proportional to the intensity of the optical mode at the lateral semiconductor-air interface in the pillar. The scattering coefficient $\kappa$ is a measure for the sidewall roughness of the micropillar and strongly depends on the applied etching process. It furthermore accounts for absorption by surface states. The third term in Eq. (2) represents material absorption in the system, and is linked to the optical absorption coefficient $\alpha$ of the semiconductor via $\alpha = \frac{2\pi n}{\lambda Q_{abs}}$ [24]. Here, $n$ represents the refractive index of the material and $\lambda$ the vacuum wavelength.

The data acquired in reflection can be most accurately reproduced using a sidewall loss coefficient of $\kappa \sim 6.8 \cdot 10^{-9} m$ and an absorption coefficient of $\alpha = 0.4 \frac{1}{cm}$. In order to reproduce the data measured in photoluminescence, we had to choose markedly higher values for both $\kappa \sim 11.2 \cdot 10^{-9} m$ and the absorption coefficient $\alpha = 0.85 \frac{1}{cm}$ which reflects the remarkable influence of the excitation conditions on the extracted Q-factor over a large diameter range. We attribute the deviation between the Q-factors extracted in µPL and reflection to a persisting photon absorption by the integrated QDs in luminescence. In µPL, the emission lines of the integrated QDs are usually subject to significant broadening mechanisms, such as pure emitter dephasing [25] as well as carrier induced broadening channels leading to spectral diffusion[26] caused by the non-resonant excitation technique. Consequently, the emission tails of slightly off resonant, however strongly broadened QDs [27] can effectively overlap with the cavity mode when this broadening occurs. In case of a not fully accomplished saturation of all QD-levels overlapping with the cavity mode, this effect can explain the increased absorption coefficient in PL experiments. We note, that due to the very high Q-factors, the excitation power in a PL experiment cannot be chosen arbitrarily high, since the systems can undergo a smooth transition into the lasing regime, where the linewidth no longer reflects the Q-factor [28,29]. In contrast, saturation of the QDs overlapping with the cavity mode in the laser reflection experiment can be established straight forwardly without risking a transition into the laser regime. A variation of the emitter density in the microcavity could further support our assumption. Such a variation could, for instance, be implemented via site-selective quantum dot positioning techniques [30]. We anticipate that carrier density dependent surface-state absorption on the micropillar sidewalls[24] can play an important role in understanding the different $\kappa$-values in µPL and µPR.

The high Q-factors of these microcavities allow us to enter the strong coupling regime of a single QD and a cavity mode of a micropillar with a diameter of 1.8 µm and a Q-factor of ~25,000 (linewidth ~ 53 µeV). Fig. 3a) shows a temperature tuning series of a single QD in close spectral vicinity of the cavity resonance. The QD was excited by a 532 nm laser under low pumping power. By modifying the sample temperature in the cryostat, the QD-emission line can be swept through the optical resonance. The strong coupling regime is manifested by the avoided crossing of the two emission peaks on resonance, as evidenced in Fig. 3b). It is worth noting that the single QD emission line is broadened up to a value of ~60 µeV, which is mainly attributed to effects of spectral diffusion. The peculiarity of our semiconductor system is that both the QD and the cavity broadening are smaller than the extracted Rabi-splitting of 66 µeV, in contrast to most other observations of strong coupling in QD-micropillars. In order to calculate the visibility of the polariton peaks in the Rabi-doublet, we have to assess the phonon induced dephasing rate of the QD-emission. We estimate this value to $\gamma^* \sim 7.8\ \mu eV$ for the QD at a temperature of ~20 K where the resonance occurs in our system [31]. From the Rabi-splitting, we can extract an interaction energy of $g$ = 35 µeV and the cavity line broadening of 53 µeV, yielding a visibility of the quantum dot-microcavity polaritons of $v = \frac{g}{\gamma^* + \gamma_c}$ ~ 0.57. With respect to the spectral diffusion induced QD broadening, this value is reduced to ~0.31, which is still markedly above the limit for the onset of strong coupling at ¼. We note, that the very large Q-factors which we observe in particular for large pillar diameters could overcompensate the reduction of the Rabi-splitting yielding even enhanced visibilities under the condition that emitter dephasing is not dominant and inhomogeneous broadenings are suppressed, which is, for instance, possible via resonant excitation techniques.

3. Conclusion

In conclusion, we have directly compared maximally extractable Q-factors in state-of the art DBR microcavities containing QD emitters. We observed a strong dependency of the cavity lineshape as well as the Q-factor on the probing technique and could directly extract record Q-factors as high as 268,000 in reflection. The extremely narrow optical resonances in these high-Q cavities allowed us to observe strong QD-cavity coupling with high visibilities. We believe that our work paves the way towards a generation of QD-micropillar devices operated in the strong coupling regime relying on distinct polariton features, such as optically or electrically driven single QD lasers in the strong coupling regime [32] or deterministic sources of indistinguishable single photons generated via the adiabatic Raman passage [35]. Furthermore, we believe that the ultra-high quality factors in conjunction with strongly coupled QD emitters, which we demonstrate in this work, will play a key role in the development of deterministic spin-photon interfaces and quantum non demolition read out schemes, as predicted in [34] and experimentally indicated in [15].

**Acknowledgments**

The authors thank M. Emmerling, A. Wolf and M. Wagenbrenner for sample preparation. We acknowledge funding the BMBF within the projects QuaHL-Rep (16BQ1042) and Q.com-H project and by the State of Bavaria.

**Figures**

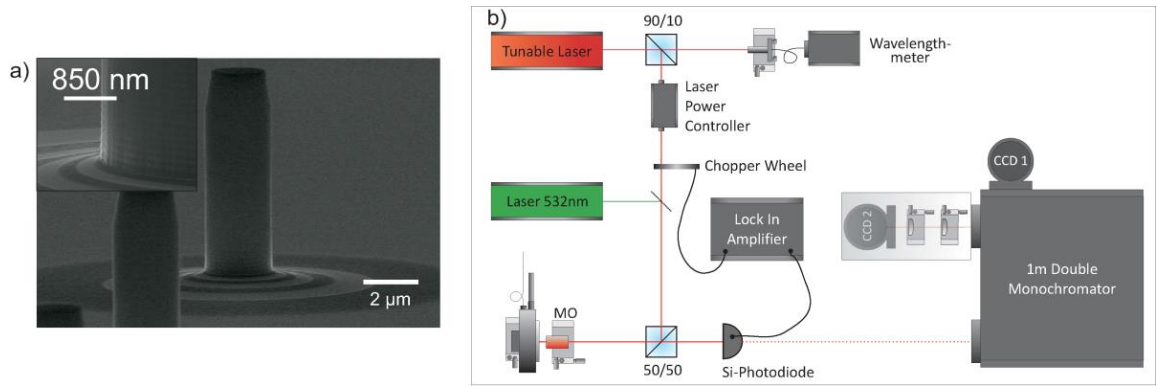

**Fig. 1:** (a) Scanning electron microscope image of a micropillar cavity with a diameter of 2.3 µm. (b) Schematic drawing of the setup which was used for photoluminescence and reflection studies.

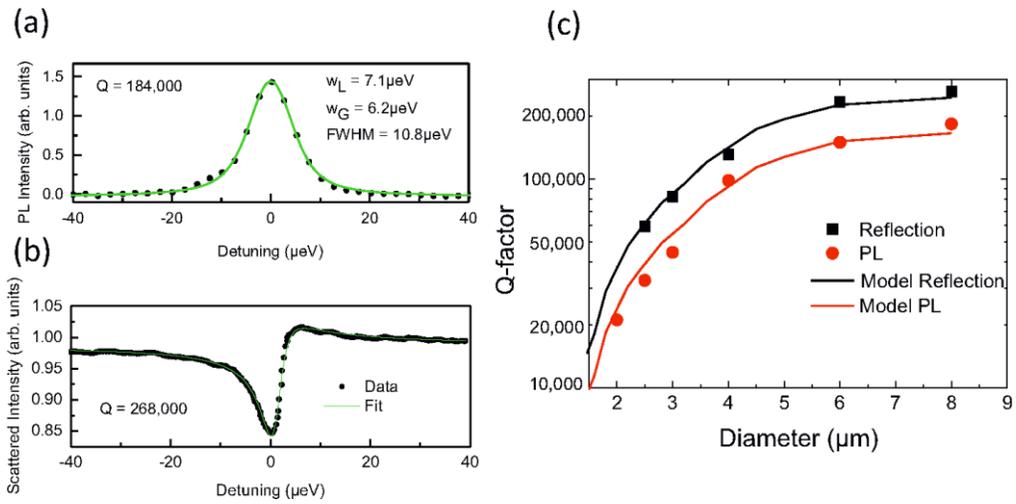

**Fig. 2:** (a) Fundamental cavity resonance from a QD-micropillar with a diameter of 6 µm measured in photoluminescence. (b) Reflectivity measurement of the same pillar, yielding a narrower resonance with a Fano-lineshape. (c) Q-factor versus pillar diameter determined in photoluminescence and reflectivity. All experiments were carried out at a sample temperature of 14 K.

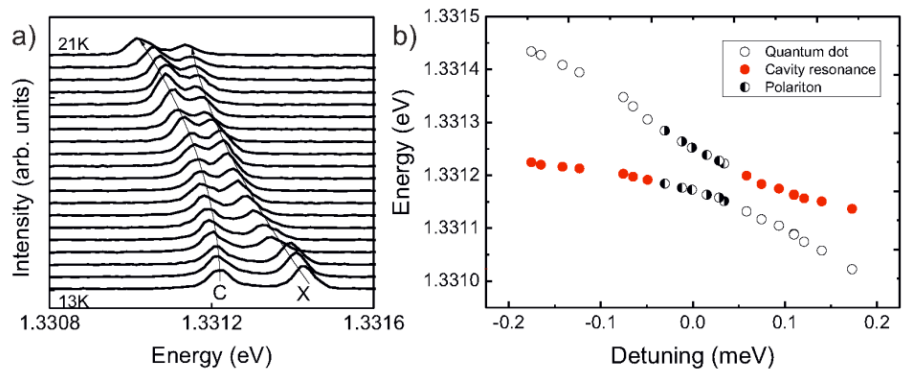

**Fig. 3:** Strong coupling of a single QD and a micropillar cavity with a Q-factor of 25,000 and a diameter of 1.8 μm. (a) Photoluminescence spectra recorded at different temperatures. The resonance case is characterized by a pronounced avoided crossing and a Rabi-splitting of 66 μeV. (b) Mode energies as a function of the QD-cavity detuning parameter.